\documentclass[12pt]{article}
\usepackage{hyperref}
\usepackage{cite}
\usepackage{color}
\usepackage{graphicx}
\usepackage{amsmath}
\usepackage{amssymb}
\usepackage{xspace}

\makeatletter
\@addtoreset{equation}{section}

\makeatletter
\renewcommand\section{\@startsection {section}{1}{\z@}%
                                   {-3.5ex \@plus -1ex \@minus -.2ex}
                                   {2.3ex \@plus.2ex}%
                                   {\normalfont\large\bfseries}}
\renewcommand\subsection{\@startsection{subsection}{2}{\z@}%
                                     {-3.25ex\@plus -1ex \@minus -.2ex}%
                                     {1.5ex \@plus .2ex}%
                                     {\normalfont\bfseries}}

\def\baselinestretch{1.2}
\newcommand{\be}{\begin{equation}}
\newcommand{\ee}{\end{equation}}
\newcommand{\beq}{\begin{eqnarray}}
\newcommand{\eeq}{\end{eqnarray}}

\newcommand{\gone}[1]{{}}

\begin{document}
\begin{titlepage}
\begin{flushright}
MAD-TH-15-02
\end{flushright}

\vfil

\begin{center}

{\bf \large
Comments on $s$-rule violating configurations in field theory}

\vfil

William Cottrell,  Akikazu Hashimoto, and Mohandas Pillai

\vfil

Department of Physics, University of Wisconsin, Madison, WI 53706, USA

\vfil

\end{center}

\begin{abstract}
\noindent We explicitly construct a configuration of ${\cal N}=4$
supersymmetry Yang-Mills theory with gauge group $U(N)$ on an interval
on length $L$ with a D5-like boundary condition on one end and an
NS5-like boundary condition on the other. For $N>1$, such a
configuration violates the $s$-rule and is non-supersymmetric. We
compute the energy relative to the BPS bound of these configurations
and find that it is proportional to $N(N^2-1) g_{YM4}^{-2} L^{-3}$.
\end{abstract}
\vspace{0.5in}

\end{titlepage}
\renewcommand{\baselinestretch}{1.05}  

\section{Introduction}

\begin{figure}[h]
\centerline{\includegraphics[scale=0.8]{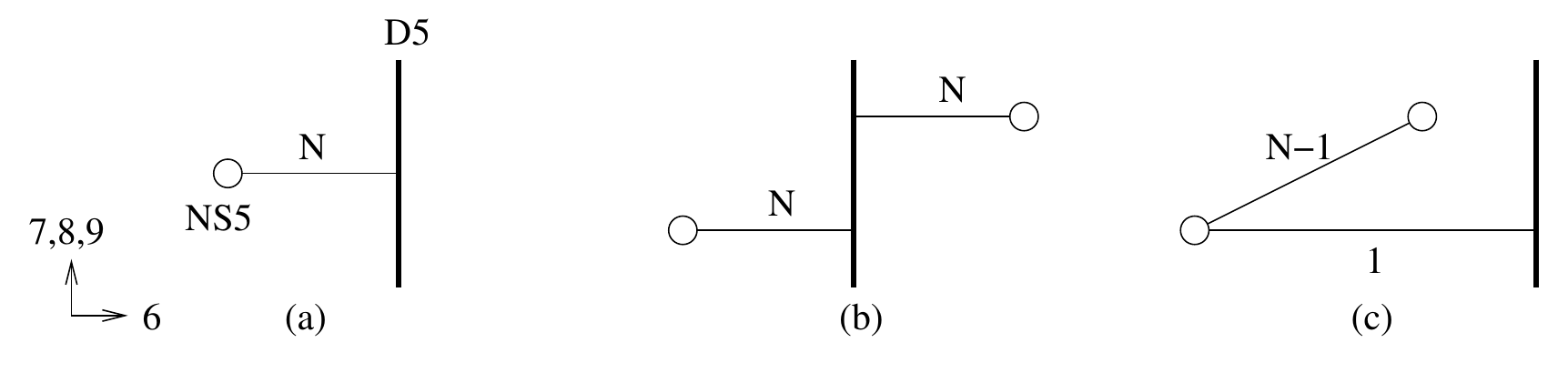}}
\caption{(a) The prototypical $s$-rule violating configuration for $N > 1$, (b) a configuration with two $s$-rule violating components, and (c) IR equivalent configuration via Hanany-Witten transition which clearly do not admit supersymmetric stationary state.  \label{figa}}
\end{figure}

In constructions involving branes in string theory, there is an
important concept known as the $s$-rule. This concept was originally
formulated in \cite{Hanany:1996ie} and states that while an arbitrary
number of D3-branes can generally end on NS5-branes and D5-branes,
when a NS5-brane and a D5-branes are oriented so that they are linked,
then not more than one D3-brane can stretch between the said NS5 and
D5-branes while preserving supersymmetry. By $\it{linked}$ we mean
that the two brane can not exchange positions by going around each
other. The setup considered in \cite{Hanany:1996ie} consisted of an
NS5-brane extended along the 012345 directions and a D5-brane extended
along the 012789 directions, separated along the $x_6$ coordinate. A
single D3-brane can stretch between the NS5 and the D5-brane. But when
two or more D3-branes are forced to stretch between these 5-branes, it
can not do so while preserving supersymmetry. The prototype
configuration violating $s$-rule is illustrated in figure \ref{figa}.a
when $N > 1$.

The argument for why this configuration breaks supersymmetry presented
originally in \cite{Hanany:1996ie} is very simple. Consider a
configuration illustrated in figure \ref{figa}.b which consists of two
$s$-rule violating components. Upon moving the D5-brane to the right
in the $x_6$ direction, the configuration turns into the one
illustrated in figure \ref{figa}.c. But the configuration in figure
\ref{figa}.c is clearly non-supersymmetric.

Although the $s$-rule seemed mysterious at first, various
reformulations that shortly followed made it much less so. For
instance, one can apply a chain of dualities to map the suspended D3
branes to a fundamental string. In such a frame, the NS5 and the D5
branes are both mapped to D-branes with 8 relatively transverse
coordinates. The massless strings stretched between D-branes oriented
that way only consists of fermions. The $s$-rule then can be viewed as
a manifestation of the Fermi exclusion principle
\cite{Danielsson:1997wq,Bachas:1997kn}. Another manifestation of the
$s$-rule can be inferred from the non-existence of supersymmetric
brane embeddings when quantum numbers of the embeddings violate the
$s$-rule. In these approaches, the dynamics of Pauli exclusion
principle is manifested classically, in the appropriate duality frame
\cite{Bachas:1997sc,Kitao:1998mf,Pelc:2000kb}. More recently, the
classical manifestation of $s$-rule was illustrated in
\cite{Hashimoto:2014nwa} in the zero slope defect field theory limit
where the NS5 and the D5-branes on which the D3 brane ends are
realized as BPS boundary conditions classified by Gaiotto and Witten
\cite{Gaiotto:2008sa,Gaiotto:2008ak}. In \cite{Hashimoto:2014nwa}, it
was argued that the Nahm pole on the D5-like boundary of ${\cal N}=4$
SYM is incompatible with the NS5-like boundary on the other end while
respecting supersymmetry.

The $s$-rule has interesting dynamical consequences. For instance, a
${\cal N} \le 3$ Chern-Simons Yang-Mills theory with gauge group
$U(N)$ and level $k$ can be engineered by suspending $N$ D3-branes
between an NS5-brane and a $(1,k)$ 5-brane\footnote{In our notion,
  $(p,q)$ 5-brane is a bound state of $p$ NS5-branes and $q$
  D5-branes.} as was considered in
\cite{Kitao:1998mf,Bergman:1999na}. Theories of this type are expected
to exhibit dynamical supersymmetry breaking when $N$ is taken to be
large. For the ${\cal N}=1$ minimal Chern-Simons Yang-Mills theory,
Witten has computed the supersymmetric index and argued that
supersymmetry is likely broken dynamically when $N > 2k$
\cite{Witten:1999ds}. For the ${\cal N}=2$ and ${\cal N}=3$ theories
arising from brane construction of \cite{Kitao:1998mf,Bergman:1999na},
Ohta has computed the Witten index and argued that supersymmetry is
likely broken for $N > k$ \cite{Ohta:1999iv}. This condition $N > k$ is
identical to the condition for the $s$-rule to be violated. This
observation supports the expectation that supersymmetry is dynamically
broken for ${\cal N}=2,3$ Chern-Simons Yang-Mills theory of
\cite{Kitao:1998mf,Bergman:1999na} in 2+1 dimensions.

Although these considerations offer considerable confidence that
dynamical supersymmetry breaking is taking place, these systems have
yet to offer intuitive understandings regarding the effective dynamics
and the scale of symmetry breaking phenomena.  One can envision taking a 't Hooft like large $N$ limit keeping $\lambda = N/k$ fixed, and one expects, on dimensional grounds, that the scale of dynamical supersymmetry breaking is parameterized as
\be \epsilon_3 = \Lambda_{DSB}^3  = \alpha (\lambda -1)^\beta (g_{YM3}^2)^3 \label{theprob}\ee
but we do not have a reliable estimate of $\alpha$ and $\beta$, nor
have we identified the effective order parameter characterizing the
supersymmetry breaking vacuum.  Some attempts to address these
questions from the field theory point of view \cite{Suyama:2012kr} as
well as using gauge/gravity correspondence
\cite{Maldacena:2001pb,Hashimoto:2010bq,Cottrell:2013asa} has so far
been inconclusive.\footnote{These papers do offer some conjectures,
  which would be interesting to confirm in an independent field theory
  analysis.}

In the classical manifestation of $s$-rule discussed from the brane
perspective in \cite{Kitao:1998mf,Bachas:1997sc,Pelc:2000kb} and from
the boundary field theory perspective \cite{Hashimoto:2014nwa}, the
absence of supersymmetric configurations violating the $s$-rule does
not preclude the existence of a non-supersymmetric configuration
solving the equation of motion and the boundary condition.  The energy
of the non-supersymmetric configuration is the vacuum energy
associated with the dynamical supersymmetry breaking, and can be
computed. This was left as an open exercise in \cite{Bachas:1997sc}.
We will first compute the profile and the energy of $s$-rule violating
configuration for a ${\cal N}=4$ SYM subjected to NS5 and D5 boundary
conditions with gauge group $U(N)$. For $N>2$, we expect the lowest
energy configuration to be non-supersymmetric.
We conclude with open issues and future directions.

\section{Boundary Field Theory Analysis}

In this section, we consider field configurations of ${\cal N}=4$
supersymmetric Yang-Mills theory in 3+1 dimensions on $R^{1,2} \times
I$, where $I$ is an interval, subjected to a D5-like boundary
condition, in the terminology of \cite{Gaiotto:2008sa,Gaiotto:2008ak}
on one end, and an NS5-like boundary condition on the other end. In
other words, this is the configuration illustrated in figure
\ref{figa}.a.

The D5-like boundary imposes a Nahm pole boundary condition. If the
boundary condition on the other end were also D5-like, this problem
reduces to the standard multi-monopole construction reviewed, for
instance, in \cite{Weinberg:2006rq}. For $N=2$, the solution takes on
the standard from involving elliptic functions. When the interval is
extended to take semi-infinite form, then one obtains the fuzzy
funnels discussed in \cite{Constable:1999ac}. If the interval is of
finite size with a D5-like boundary condition on one end and impose
the $N$ NS5-like boundary condition on the other, then there is a BPS
configuration which was worked out in section 3.6 of
\cite{Hashimoto:2014nwa}. 

When the $U(N)$ on an interval is forced to respect D5-like boundary
condition on one end but NS5-like boundary on the other, there is a
tension between the Nahm pole blow up along the 789 coordinates along
with the D5 is extended while the NS5 is localized and imposes a
Dirichlet boundary condition. Because of this tension, solutions to the first order BPS equation with these boundary conditions are 
impossible unless $N=1$.

One could, however, look for a non-BPS solution to the second order
equation of motion. Consider the bosonic component of ${\cal N}=4$ supersymmetric Yang-Mills theory in 3+1 dimensions viewed as a dimensional reduction of ${\cal N}=1$ supersymmetric Yang-Mills theory in 9+1 dimensions.

The action of this theory can be written simply as
\be S  = -{1 \over 4 g_{YM4}^2} \int \mbox{Tr} F_{ij}F^{ij} \ee
where
\be F_{ij} = \partial_i A_j - \partial_j A_i + i [A_i,A_j] \ee
and
\be g_{YM4}^2 = 2 \pi g_s \ee 
following the standard conventions in string theory (see e.g (212) and
(275) of \cite{Johnson:2000ch}). Note that it is somewhat
unconventional to normalize the non-abelian gauge kinetic term in the
trace form with a factor of $1/4$. In order to relate to the standard
convention used e.g. in \cite{Weinberg:2006rq} where the action is
presented as
\be S = -{1 \over 2 e^2} \int \mbox{Tr} F_{ij}F^{ij} \ ,
\ee 
one must relate
\be e^2 = 2 g_{YM4}^2  \ . \ee
For static configurations, the energy density
\be  \epsilon_4 = - {\cal L} = {1 \over 4 g_{YM4}^2}  \mbox{Tr} F_{ij}F^{ij} \ee

The ansatz we consider is extremely simple, namely;
\be A_{6+i} = f(z) T^i, \qquad i = 1\ldots 3 \label{ansatz}\ee
where $z$ parameterizes the $x_6$ coordinate, and with other $A_i$ set
to zero, and $T^i$ are the $N$ dimensional generators of $SU(2)$. For
$N=2$,
\be T^i = {1 \over 2} \sigma^i, \ee
and for general $N$, 
\be \mbox{Tr} \sum_i (T^i)^2 = {N(N^2-1)  \over 4} \ .  \label{trace} \ee
The configuration we seek is a solution to the equations of motion with the boundary condition that at $z=0$, the solution approaches the Nahm pole \cite{Gaiotto:2008sa,Gaiotto:2008ak}
\be f(z) = {1 \over z} \label{nahmpole} \ee
with coefficient $1$ and at $z=L$ for some fixed $L$, 
\be f(z) = 0 \ee
to respect the Dirichlet boundary condition imposed by the NS5-brane.

Upon substituting the ansatz (\ref{ansatz}) to the Yang-Mills equation of motion, we simply obtain an equation of for $f(z)$ which reads
\be f''(z) - 2 f^3(z)=0  \ee
This is essentially the equation of motion for $\phi^4$ theory dimensionally reduced to 0+1 dimension. It can also be viewed as the equation of motion for a non-linear spring.

This equation of motion can also be written in the form
\be {(f'(z)^2 - f(z)^4)' \over 2 f'(z)} = 0 \ . \ee 
which implies
\be f'(z)^2 - f(z)^4 = c \label{conserved} \ee
is conserved, and this equation can further be integrated 
\be  -{df \over \sqrt{c + f^4}} = dz \ee
The two integration constants cay be fixed by requiring $f=0$ at $z=L$ and $f=\infty$ at $z=0$, i.e. 
\be 
L = \int_0^\infty {df \over \sqrt{c + f^4}}  = {1 \over 4 \sqrt{\pi}} c^{-1/4} \Gamma\left(\tfrac{1}{4}\right)^2
\ee
from which we read off that
\be c = {\Gamma\left(\tfrac{1}{4} \right)^8 \over 256 \pi^2 L^4} \ . \ee
It is clear, then, that near $z=0$, the Nahm pole boundary condition (\ref{nahmpole}) is satisfied. 

\begin{figure}
\centerline{\includegraphics[scale=0.8]{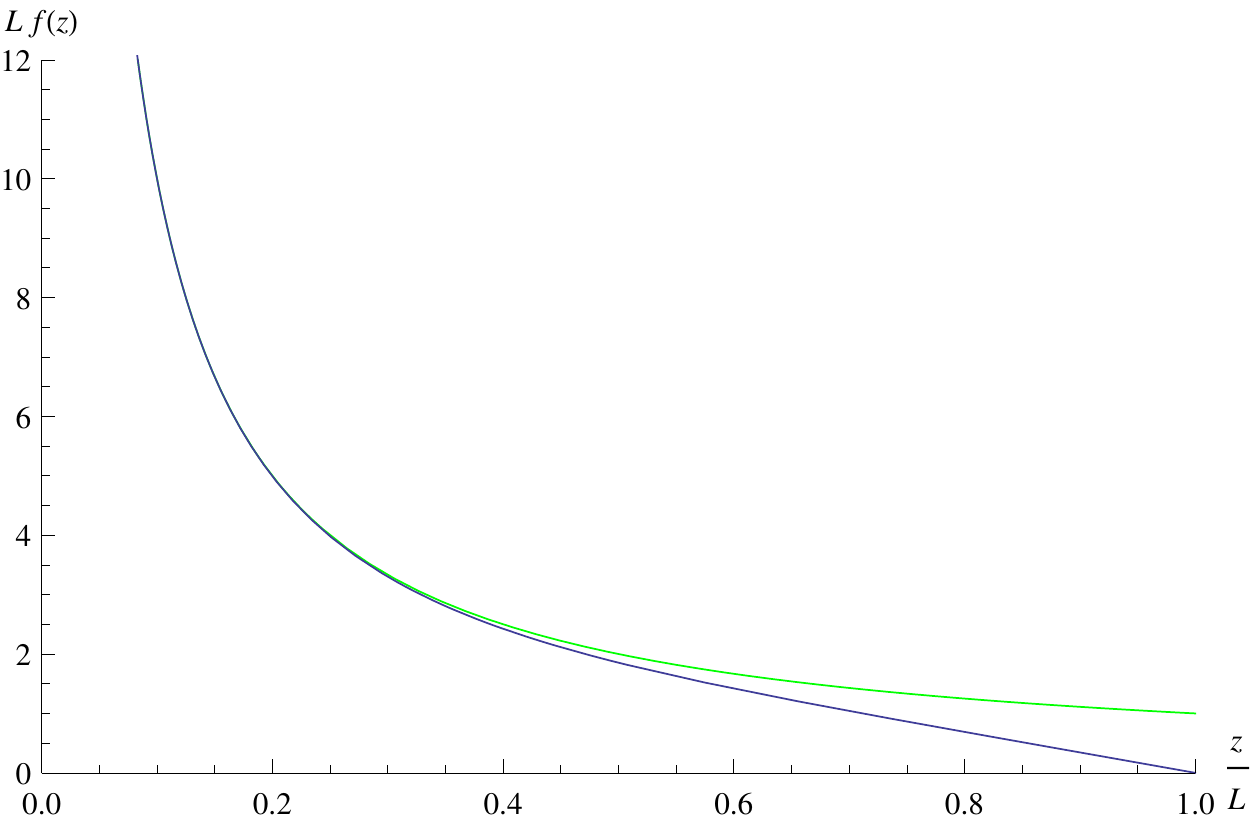}}
\caption{The solution $f(z)$ which reflects the profile of a
  non-abelian funnel-like structure for the $s$-rule violating
  configuration of $U(N)$ gauge theory on an interval $0 < z < L$ with
  a D5-like boundary at $z=0$ and an NS5-like boundary at $z=L$ is illustrated by the blue curve. The green curve is the BPS solution corresponding to the Nahm pole/fuzzy funnel $f(z)=1/z$. 
\label{figb}}
\end{figure}

The full solution consistent with these boundary conditions can be expressed in terms of a hyper-geometric function 
\be z  = { {}_2 F_1\left(\tfrac{1}{4}, \tfrac{1}{2}, \tfrac{5}{4}, -{\Gamma({1 \over 4})^8 \over 256 \pi^2 L^4 f^4}\right)  \over f}  \ee
and has the form illustrated in figure \ref{figb}. Note that the solution asymptotes to $f=1/z$ near $z=0$ while approaching $f=0$ at $z=L$. The only exception is the case of $N=1$ for which $T^i=0$ and therefore the Nahm pole is absent and $f=0$ is the trivial, BPS, solution.

Having found the stationary field configuration associated with a non-BPS state, it would be interesting to compute its energy. Substituting the ansatz (\ref{ansatz}) in to the energy density, we find
\be \epsilon_4 = {1 \over 2 g_{YM4}^2} \mbox{Tr} \sum_i (T^i)^2 \left(f'^2+f^4 \right) 
={N(N^2-1) \over 8 g_{YM4}^2}  \left(c+ 2 f^4 \right) \ee
having used (\ref{conserved}) and (\ref{trace}).

An interesting quantity is the effective three dimensional energy density obtained by integrating 
\be \epsilon_3 = {N(N^2-1) \over 8 g_{YM4}^2} \int_0^L dz \, (c + 2 f^4)
=  {N(N^2-1) \over 8 g_{YM4}^2} \int_0^\infty df \, {c + 2 f^4 \over \sqrt{c+f^4}} \label{e3}
\ee
which diverges as $z$ approaches $0$ where $f$ goes to infinity.

We should recall, however, that the quantity of interest is the energy above the BPS bound.  The energy density can be written in the form
\be \epsilon_4 = \epsilon_4^{non-BPS} + \epsilon_{4}^{BPS} \ee
where
\be \epsilon_4^{non-BPS}  = {1 \over 2 g_{YM4}^2} \mbox{Tr} \left({d A^i \over dz} + {i \over 2} \epsilon_{ijk} [A_j,A_k]\right)^2 \ee
is the positive definite non-extremal contribution, and
\be \epsilon_4^{BPS} = -{i \over 3 g_{YM4}^2} {d \over dz} \mbox{Tr} (\epsilon_{ijk} A_i A_j A_k) \ee
is the contribution which one expects from a BPS configuration. To extract the non-extremal component, we should add $\epsilon_3^{BPS}$ for our ansatz, which takes the form
\be \epsilon_3^{BPS} = \int_0^z dz \epsilon_4^{BPS} = -{N(N^2-1) \over 8 g_{YM4}^2} \int_0^\infty df\, 2f^2 \ , \ee 
to (\ref{e3}). Similar consideration of separating the BPS component
from the non-extremal part was discussed in (9) of
\cite{Constable:2001ag}. The non-extremal contribution to the energy
inferred this way is 
\be \epsilon_3^{non-BPS} = 
 {N(N^2-1) \over 8 g_{YM4}^2} \int_0^\infty df\,\left( \sqrt{c+2 f^4} - 2 f^2 \right)  = \#  {N(N^2-1) \over g_{YM4}^2 L^3}  \label{mainresult}
\ee
where $\#$ is a numerical factor of order one which is easily calculable. 

Equation (\ref{mainresult}) is the main result of this paper. It is
the leading small $g_{YM4}$ behavior of the non-extremal contribution
to the energy of ${\cal N}=4$ $U(N)$ supersymmetric Yang-Mills theory
on an interval of length $L$ with a D5-like boundary condition on one
end and an NS5-like boundary condition on the other. Although the
equations of motion and the boundary condition respects half of the
supersymmetries of the ${\cal N}=4$ theory, the stationary solution is
not supersymmetric. As such, this is an example of spontaneously
broken supersymmetry.  The dependence on $g_{YM4}^2$ and $L^3$ may
have been anticipated from dimensional grounds, but the dependence on
$N$ is somewhat non-trivial. The exercise of computing the
non-extremal energy for these non-supersymmetric stationary states was
suggested, for instance, at the end of \cite{Bachas:1997sc}. In this
paper, we reported on a simple ansatz which allowed this exercise to
be carried out in a closed form by working in a context where $s$-rule
is manifested in a strictly field theoretic, zero slope limit of
string theory \cite{Hashimoto:2014nwa}.

\section{Discussion}

The system we considered in this note, namely ${\cal N}=4$
supersymmetric $U(N)$ Yang-Mills theory on an interval with NS5-like
and D5-like boundaries on each ends, is a simple example of a theory
exhibiting dynamical supersymmetry breaking in that the equation of
motion and boundary conditions are supersymmetric but the solution to
the equations are not. Being defined on an interval, the theory is
effectively 2+1 dimensions at long distances, but the dynamics of
supersymmetry breaking relied on full 3+1 dimensional physics. A case
in point is that the scale of supersymmetry breaking scales like
$L^{-1}$ and diverges in the small $L$ limit. Also, this system is
empty in the deep IR limit and does not have a clean interpretation as
a 2+1 dimensional system in the first place.

There are several generalizations to our exercise that one can
consider. One which immediately comes to mind is to generalize the
NS5-like boundary condition corresponding to the single NS5, to that
of a stack of $k$ NS5-branes. Then, the $s$-rule will permit, as was
demonstrated in \cite{Hashimoto:2014nwa}, up to $N=k$ D3-branes with a
Nahm pole on a D5-like boundary on the other end. Once the boundary
condition are generalized this way, it is likely that multiple,
possibly a continuous family, of solutions exist for a given
quantum number $N$. It would be interesting to enumerate these
possibilities explicitly.

A setup that would be extremely interesting to understand is the
energy of $s$-rule violating configuration of $N$ D3-branes stretched
between a NS5-like boundary and a $(1,k)$ 5-brane-like boundary on the
other, oriented in such a way as to engineer ${\cal N}=2$ or ${\cal N}=3$
$U(N)$ Chern-Simons Yang-Mills theory in 2+1 dimensions with level $k$
\cite{Bergman:1999na,Kitao:1998mf}. Unfortunately, for this setup, it
appears that one must analyze the quantum effects to demonstrate the
spontaneous breaking of supersymmetry.

The dynamics of spontaneous supersymmetry breaking will manifest itself
in the S-dual system consisting of a D5-like boundary on one end and a
$(k,1)$-like boundary on the other. Unfortunately, as discussed in
section 8.3 of \cite{Gaiotto:2008ak}, the $(k,1)$-like boundary
appears to be somewhat subtle, and have neither a concrete
understanding of the BPS configuration for $N \le k$ nor of the
non-BPS configurations with $N > k$.  Even if we did manage to
understand the manifestation of dynamical supersymmetry breaking
classically in this duality frame, it will not be quantitatively
reliable in the limit where the 2+1 dimensional Yang-Mills coupling is
taken to be much smaller than the scale of the interval $g_{YM2}^2 \ll
L^{-1}$, so some other approach would be required to address the
problem of computing $\alpha$ and $\beta$ in (\ref{theprob}). 

Another possible extension of our work is to study the
nonsupersymmetric $s$-rule violating brane embeddings which was posed
in the conclusion of \cite{Bachas:1997sc}. The BPS embeddings in
various manifestations of the configurations respecting the $s$-rule
have been presented in the literature
\cite{Bachas:1997sc,Pelc:2000kb}. Unfortunately, the embedding appears
to be rather complicated even for the first order BPS equations, and
it is not immediately clear how one can extend this exercise to solve
the full second order equation of motion. 
We hope to present better understating of these issues in the near future \cite{inprogress}.

\section*{Acknowledgements}

This work supported in part by funds from University of Wisconsin.  AH
would like to thank Peter Ouyang and Masahito Yamazaki for
collaboration on related work which inspired this project.

\bibliography{srule}\bibliographystyle{utphys}

\providecommand{\href}[2]{#2}\begingroup\raggedright\begin{thebibliography}{10}

\bibitem{Hanany:1996ie}
A.~Hanany and E.~Witten, ``{Type IIB superstrings, BPS monopoles, and
  three-dimensional gauge dynamics},'' {\em Nucl.Phys.} {\bf B492} (1997)
  152--190,
\href{http://www.arXiv.org/abs/hep-th/9611230}{{\tt hep-th/9611230}}.

\bibitem{Danielsson:1997wq}
U.~Danielsson, G.~Ferretti, and I.~R. Klebanov, ``{Creation of fundamental
  strings by crossing D-branes},'' {\em Phys.Rev.Lett.} {\bf 79} (1997)
  1984--1987,
\href{http://www.arXiv.org/abs/hep-th/9705084}{{\tt hep-th/9705084}}.

\bibitem{Bachas:1997kn}
C.~P. Bachas, M.~B. Green, and A.~Schwimmer, ``{(8,0) quantum mechanics and
  symmetry enhancement in type I' superstrings},'' {\em JHEP} {\bf 9801} (1998)
  006,
\href{http://www.arXiv.org/abs/hep-th/9712086}{{\tt hep-th/9712086}}.

\bibitem{Bachas:1997sc}
C.~Bachas and M.~B. Green, ``{A classical manifestation of the Pauli exclusion
  principle},'' {\em JHEP} {\bf 9801} (1998) 015,
\href{http://www.arXiv.org/abs/hep-th/9712187}{{\tt hep-th/9712187}}.

\bibitem{Kitao:1998mf}
T.~Kitao, K.~Ohta, and N.~Ohta, ``{Three-dimensional gauge dynamics from brane
  configurations with $(p,q)$ - five-brane},'' {\em Nucl.Phys.} {\bf B539}
  (1999) 79--106,
\href{http://www.arXiv.org/abs/hep-th/9808111}{{\tt hep-th/9808111}}.

\bibitem{Pelc:2000kb}
O.~Pelc, ``{On the quantization constraints for a D3-brane in the geometry of
  NS5-branes},'' {\em JHEP} {\bf 0008} (2000) 030,
\href{http://www.arXiv.org/abs/hep-th/0007100}{{\tt hep-th/0007100}}.

\bibitem{Hashimoto:2014nwa}
A.~Hashimoto, P.~Ouyang, and M.~Yamazaki, ``{Boundaries and defects of $
  \mathcal{N}=4 $ SYM with 4 supercharges. Part II: Brane constructions and 3d
  $ \mathcal{N}=2 $ field theories},'' {\em JHEP} {\bf 1410} (2014) 108,
\href{http://www.arXiv.org/abs/1406.5501}{{\tt 1406.5501}}.

\bibitem{Gaiotto:2008sa}
D.~Gaiotto and E.~Witten, ``{Supersymmetric boundary conditions in ${\cal N}=4$
  super Yang-Mills theory},'' {\em J.Statist.Phys.} {\bf 135} (2009) 789--855,
\href{http://www.arXiv.org/abs/0804.2902}{{\tt 0804.2902}}.

\bibitem{Gaiotto:2008ak}
D.~Gaiotto and E.~Witten, ``{S-duality of boundary conditions in ${\cal N}=4$
  super Yang-Mills theory},'' {\em Adv.Theor.Math.Phys.} {\bf 13} (2009) 721,
\href{http://www.arXiv.org/abs/0807.3720}{{\tt 0807.3720}}.

\bibitem{Bergman:1999na}
O.~Bergman, A.~Hanany, A.~Karch, and B.~Kol, ``{Branes and supersymmetry
  breaking in three-dimensional gauge theories},'' {\em JHEP} {\bf 9910} (1999)
  036,
\href{http://www.arXiv.org/abs/hep-th/9908075}{{\tt hep-th/9908075}}.

\bibitem{Witten:1999ds}
E.~Witten, ``{Supersymmetric index of three-dimensional gauge theory},''
\href{http://www.arXiv.org/abs/hep-th/9903005}{{\tt hep-th/9903005}}.

\bibitem{Ohta:1999iv}
K.~Ohta, ``{Supersymmetric index and $s$-rule for type IIB branes},'' {\em
  JHEP} {\bf 9910} (1999) 006,
\href{http://www.arXiv.org/abs/hep-th/9908120}{{\tt hep-th/9908120}}.

\bibitem{Suyama:2012kr}
T.~Suyama, ``{Supersymmetry Breaking in Chern-Simons-matter Theories},'' {\em
  JHEP} {\bf 1207} (2012) 008,
\href{http://www.arXiv.org/abs/1203.2039}{{\tt 1203.2039}}.

\bibitem{Maldacena:2001pb}
J.~M. Maldacena and H.~S. Nastase, ``{The supergravity dual of a theory with
  dynamical supersymmetry breaking},'' {\em JHEP} {\bf 0109} (2001) 024,
\href{http://www.arXiv.org/abs/hep-th/0105049}{{\tt hep-th/0105049}}.

\bibitem{Hashimoto:2010bq}
A.~Hashimoto, S.~Hirano, and P.~Ouyang, ``{Branes and fluxes in special
  holonomy manifolds and cascading field theories},'' {\em JHEP} {\bf 1106}
  (2011) 101,
\href{http://www.arXiv.org/abs/1004.0903}{{\tt 1004.0903}}.

\bibitem{Cottrell:2013asa}
W.~Cottrell, J.~Gaillard, and A.~Hashimoto, ``{Gravity dual of dynamically
  broken supersymmetry},'' {\em JHEP} {\bf 1308} (2013) 105,
\href{http://www.arXiv.org/abs/1303.2634}{{\tt 1303.2634}}.

\bibitem{Weinberg:2006rq}
E.~J. Weinberg and P.~Yi, ``{Magnetic Monopole Dynamics, Supersymmetry, and
  Duality},'' {\em Phys.Rept.} {\bf 438} (2007) 65--236,
\href{http://www.arXiv.org/abs/hep-th/0609055}{{\tt hep-th/0609055}}.

\bibitem{Constable:1999ac}
N.~R. Constable, R.~C. Myers, and O.~Tafjord, ``{The Noncommutative bion
  core},'' {\em Phys.Rev.} {\bf D61} (2000) 106009,
\href{http://www.arXiv.org/abs/hep-th/9911136}{{\tt hep-th/9911136}}.

\bibitem{Johnson:2000ch}
C.~V. Johnson, ``{D-brane primer},''
\href{http://www.arXiv.org/abs/hep-th/0007170}{{\tt hep-th/0007170}}.

\bibitem{Constable:2001ag}
N.~R. Constable, R.~C. Myers, and O.~Tafjord, ``{Non-abelian brane
  intersections},'' {\em JHEP} {\bf 0106} (2001) 023,
\href{http://www.arXiv.org/abs/hep-th/0102080}{{\tt hep-th/0102080}}.

\bibitem{inprogress}
 Work in progress.

\end{thebibliography}\endgroup

\end{document}